\documentclass{acm_proc_article-sp}
\usepackage[utf8]{inputenc}
\usepackage{tabularx}
\usepackage{longtable}
\usepackage{url}
\usepackage{balance}
\usepackage{microtype}
\usepackage[caption=false]{subfig}

\begin{document}

\title{Identifying And Weighting Integration Hypotheses On Open Data Platforms}

\numberofauthors{1}
\author{\alignauthor Julian Eberius, Katrin Braunschweig, Maik Thiele and Wolfgang Lehner\\
\affaddr{Technische Universit\"at Dresden\\
Faculty of Computer Science, Database Technology Group
}\\
\affaddr{01062 Dresden, Germany} \\
\email{firstname.lastname@tu-dresden.de}}

\maketitle

\begin{abstract}
Open data platforms such as \url{data.gov} or \url{opendata.socrata.com} provide a huge amount of valuable information.
Their free-for-all nature, the lack of publishing standards and the multitude of domains and authors represented on these platforms lead to new integration and standardization problems.
At the same time, crowd-based data integration techniques are emerging as new way of dealing with these problems.
However, these methods still require input in form of specific questions or tasks that can be passed to the crowd.
This paper discusses integration problems on Open Data Platforms, and proposes a method for identifying and ranking integration hypotheses in this context. We will evaluate our findings by conducting a comprehensive evaluation using on one of the largest Open Data platforms.
\end{abstract}




\section{Open Data and Reuseability} 
\label{sec:open_data_and_reuseability}
Following the \emph{Open Data} trend, governments and public agencies have started to make their data available to the public using web portals, web services or REST interfaces.
One of the central ideas of the Open Data movement is that public availability of as many datasets as possible will allow reuse of the data in new and unforeseen circumstances.
Ideally, this would drive innovation as much as it would lead to a more democratic and transparent society.
In this paper we treat so called \emph{Open Data Platforms}, such as \url{data.gov}, or \url{data.un.org}, publishing platforms that contain heterogeneous, mostly unconnected datasets from disparate sources.
In particular, we do not assume that the data is organized as Linked Data, but as disjoint datasets, as this is much more commonly encountered on existing Open Data Platforms.\\
To achieve the goal of easy reusability, datasets on these platforms should be as clean, standardized, and integrated as possible.
This could mean using common vocabulary, standardized metadata or at best even integrated global schemata that are shared between datasets.
For example, if two datasets reference the same real world entities, this fact should be made explicit in metadata and should be queryable.\\
A second ideal of Open Data is that as much data as possible should be published, and that the broadest possible spectrum of contributers should be included in the publication process.
For example, many of the existing platforms are free-for-all, meaning that all users can contribute datasets.
And even on those that have some central coordination, like the big government platforms, many very different agencies and individual civil servants post datasets.
It is immediately apparent that the mentioned goals, easy reusability and strong integration on the one side, versus large quantity of datasets and large number of contributors on the other side, are contradictory.
This is supported by studies into existing Open Data Platforms, e.g., in \cite{Braunschweig:2012}.
Since limiting publication or imposing too many rules on publishers is not in the spirit of Open Data, new approaches to data integration will be necessary to cope with the problems of free-for-all data platforms.\\
The next section will describe where we see the specific differences in data integration on Open Data Platforms when compared to classical integration scenarios.
We will argue why the identification and weighting of integration problems is a necessary preliminary stage to the actual integration in these new environments.
We will also describe novel classes of integration problems, so called \emph{global integration problems}, that did not occur in classical integration scenarios.
Section \ref{sec:a_global_datamodel_for_open_data_platforms} will introduce our data model which we use to perform the integration preprocessing. Section \ref{sec:identifying_and_weighting_global_integration_problems} will describe the methods we use for problem identification and our weighting schemes, while Section \ref{sec:experimental_evaluation} will evaluate our method on an existing Open Data platform, \emph{opendata.socrata.com}.
We will discuss related word in Section \ref{sec:related_work} and give directions for future work in Section \ref{sec:conclusion_and_future_work}.

\section{Challenges in Data Integration On Open Data Platforms} 
\label{sec:challenges_in_data_integration}
In this section we introduce two factors in data integration and their occurrence on Open Data Platforms and discuss how they motivate our approach.
\subsection{Global Data Integration} 
\label{sub:global_data_integration}
Traditional data integration is goal oriented: integration tasks are specified for a defined set of data sources and defined application or analysis goal.
In contrast, data on Open Data Platforms would have to be integrated for the sake of having integrated data, as the possible reuse scenarios are not known beforehand.
Furthermore, in classical data integration is usually considered as one-to-one integration between two well-defined schemata, i.e. sets of relations, that describe the same domain.
On an Open Data Platform, there is a large number of mostly unrelated datasets that usually have very few corresponding attributes if at all.
Still, subsets of them describe the same domains and could be reused together if they were properly consolidated.
So while there is reuse and recombination potential, making all the different datasets obey to a global schema is unfeasible.
This view of Open Data Platforms is similar to the concept of dataspaces \cite{Franklin:2005}.
The current philosophy in working with such dataspaces is the so called pay-as-you-go approach\cite{Madhavan:2007}, in which integration is postponed until it is clear how the data should be reused.
Still, we argue that some integration tasks can and should be tackled a priori, to improve the usefulness of Open Data Platforms as a whole.
For example, performing as much standardization as possible before reuse, e.g. using common terms in datasets describing the same domain, will make datasets as well potentially matching sets more discoverable.
Other examples would be the mentioned problems of annotating datasets with standardized temporal or spatial metadata or connecting datasets dealing with the same entities.\\
While existing schema- and instance matching techniques can be applied to perform these integration tasks, one additional challenge in this scenario is to identify the combinations of datasets that have integration potential.
It is unfeasible to manually inspect all possible combinations of datasets for their mutual integration potential. Since there is no global schema, comparing each individual dataset to a central schema is not an option either.
We argue that identifying integration problems between datasets that can be solved traditionally requires a global view of all available datasets.
In addition to global integration potential analysis, Open Data Platforms have some unique integration problems that do not appear in classical integration scenarios and which can only be identified using a global view on the level of datasets.
These problems include partial- or duplicated datasets, partitioned datasets, versioned datasets and others, which will be described in detail in Section \ref{sec:identifying_and_weighting_global_integration_problems}.

\subsection{Crowd-based Data Integration} 
\label{sub:crowdsourcing_data_integration}
Apart from these challenges, there is also the factor of integration costs.
In general, data integration is very expensive, as it is a laborious task that has to be performed by highly trained experts.
As the number of public datasets is high and will likely grow, while the commercial interest for paying the integration will not always be clear a priori, it is unrealistic to assume that experts will be available to solve integration problems.
A solution that is currently gaining traction in the data integration community is using crowdsourcing approaches to solve tasks that can not be solved using automatic methods, e.g. in \cite{Franklin:2011}, and \cite{McCann:2008}.
While these approaches can partially substitute expert work in data integration, they pose the new challenge of deciding which questions should be answered by the crowd.
Specifically, a decision has to be made whether the output of an automatic integration algorithm can be relied on, or whether the result should be verified by humans.
To give an example: Two datasets on an Open Data Platform are deemed to be concerned with the same domain, because their attribute are found to be correspondent by a schema matching algorithm.
If all attributes match exactly, human intervention might not be necessary.
If, on the other hand, only a subset of attributes matches or the matching algorithm returns low confidence values, it might be a good next step to verify this result using the crowd.
If instance data is taken into account, the exemplary problems becomes even more intricate: on an Open Data Platform two datasets can have a strong overlap in instances, but have completely different metadata.
As we will see in section \ref{sec:identifying_and_weighting_global_integration_problems}, this could be a hint that the same dataset was uploaded by different users with different metadata.
Again, deciding whether this is really the case, and whether the two datasets should be reconciled is hard for automatic algorithms.
Even though there is some potential for volunteer work when the datasets to be integrated are of great public interest, simply validating all results of automatic methods is not feasible.
Therefore, we argue that is not only necessary to automatically identify integration potentials on Open Data Platforms, it is also necessary to weight them, for example according to difficulty or importance, to direct the work of the crowd to the most pressing integration problems.

To summarize, assuming the two new necessities of \emph{global data integration} and \emph{directing crowdsourcing in data integration} we have come to the conclusion that a preprocessing step to the actual data integration is necessary.
Specifically, we argue that it is necessary to identify potential integration problems in big sets of independent datasets, and that schemes for weighting the relevance of these problems is necessary.

\section{A Global Content Model for\\ Open Data Platforms} 
\label{sec:a_global_datamodel_for_open_data_platforms}
\begin{figure}[t] \centering
\includegraphics[width=0.35\textwidth]{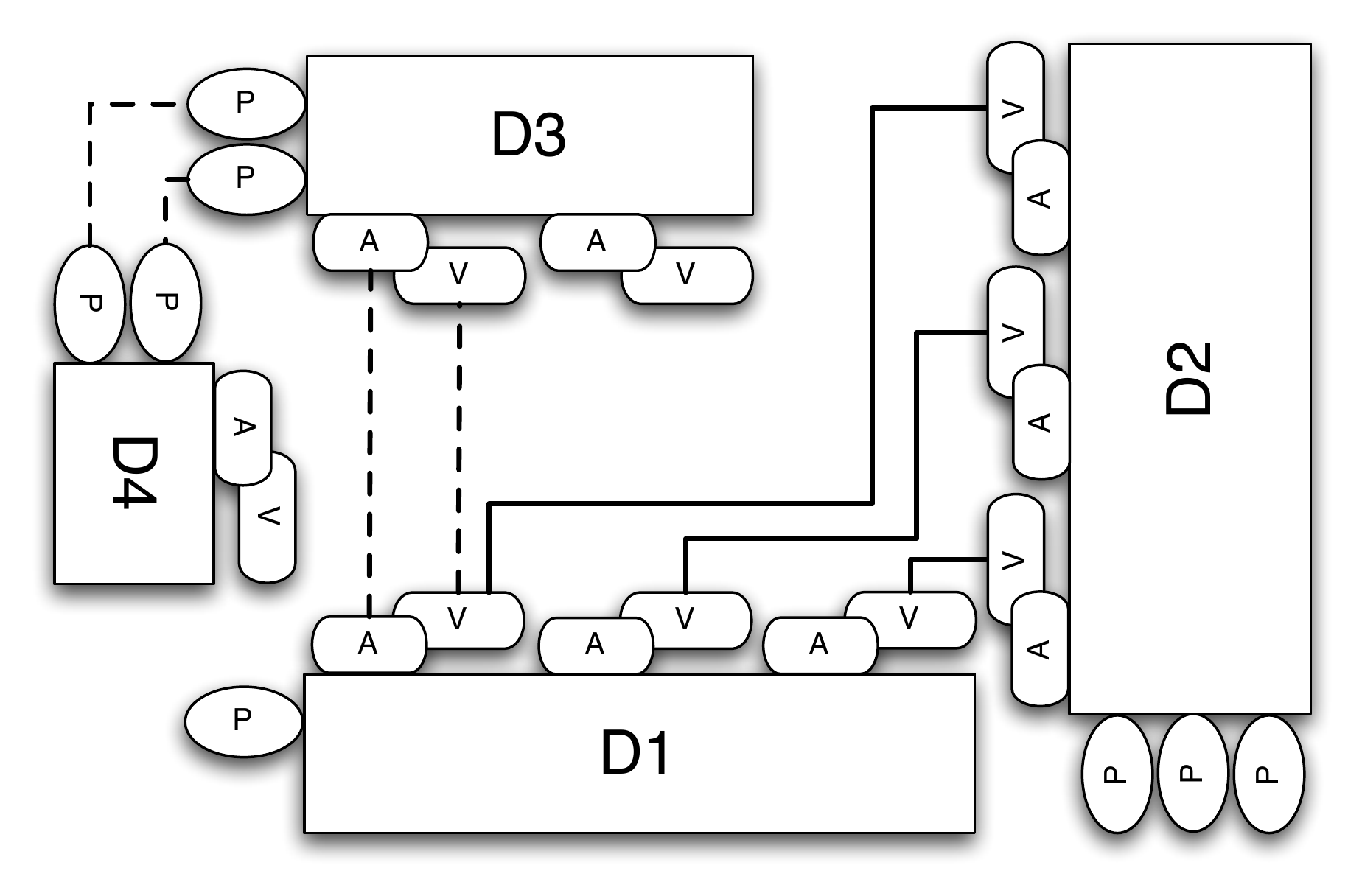}
\caption{Example for a global content model for a platform with four datasets. Different line strokes represent different edge scores.}
\label{fig:ex0}
\end{figure}
By studying existing Open Data Platforms and reducing them to their smallest set of common properties we devised a simple model that represents the published datasets as well as their manifold relationships.
It can be instantiated for a given platform and forms the basis for our problem classification, as well as the algorithms to identify and weight these problems.
In this model, a platform's content is modeled as a set $D$ of datasets with each dataset $d$ consisting of attributes, value sets for each attribute, and generic metadata properties: $d = (A, V, P)$.
This is the factual part of the model.
\begin{figure*}[t]
  \centering
  \subfloat[Duplicated Dataset (Re-upload)]{\label{fig:ex1}\includegraphics[width=0.3\textwidth]{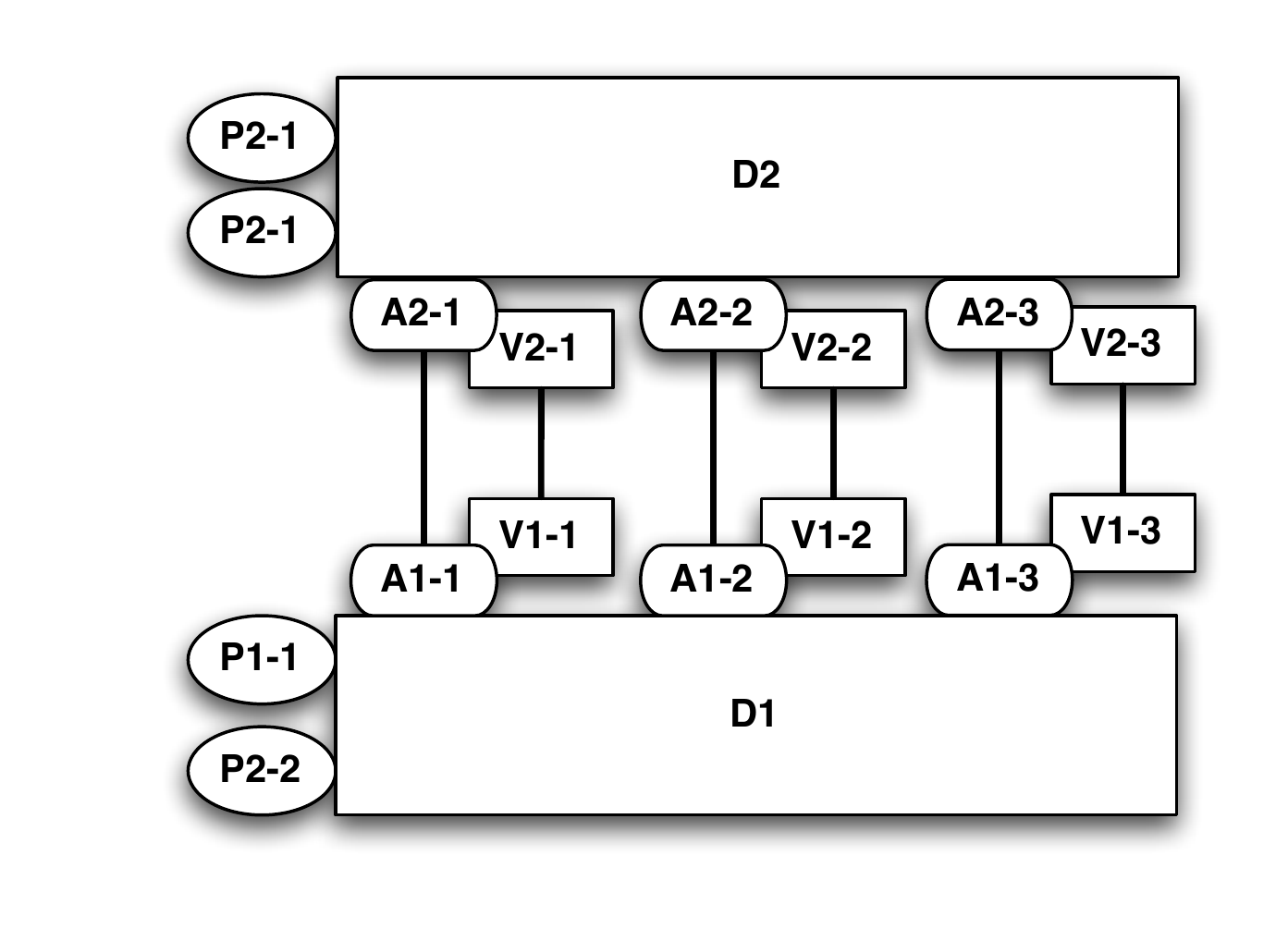}}
  \subfloat[Sub-/Supersets or Versions]{\label{fig:ex3}\includegraphics[width=0.3\textwidth]{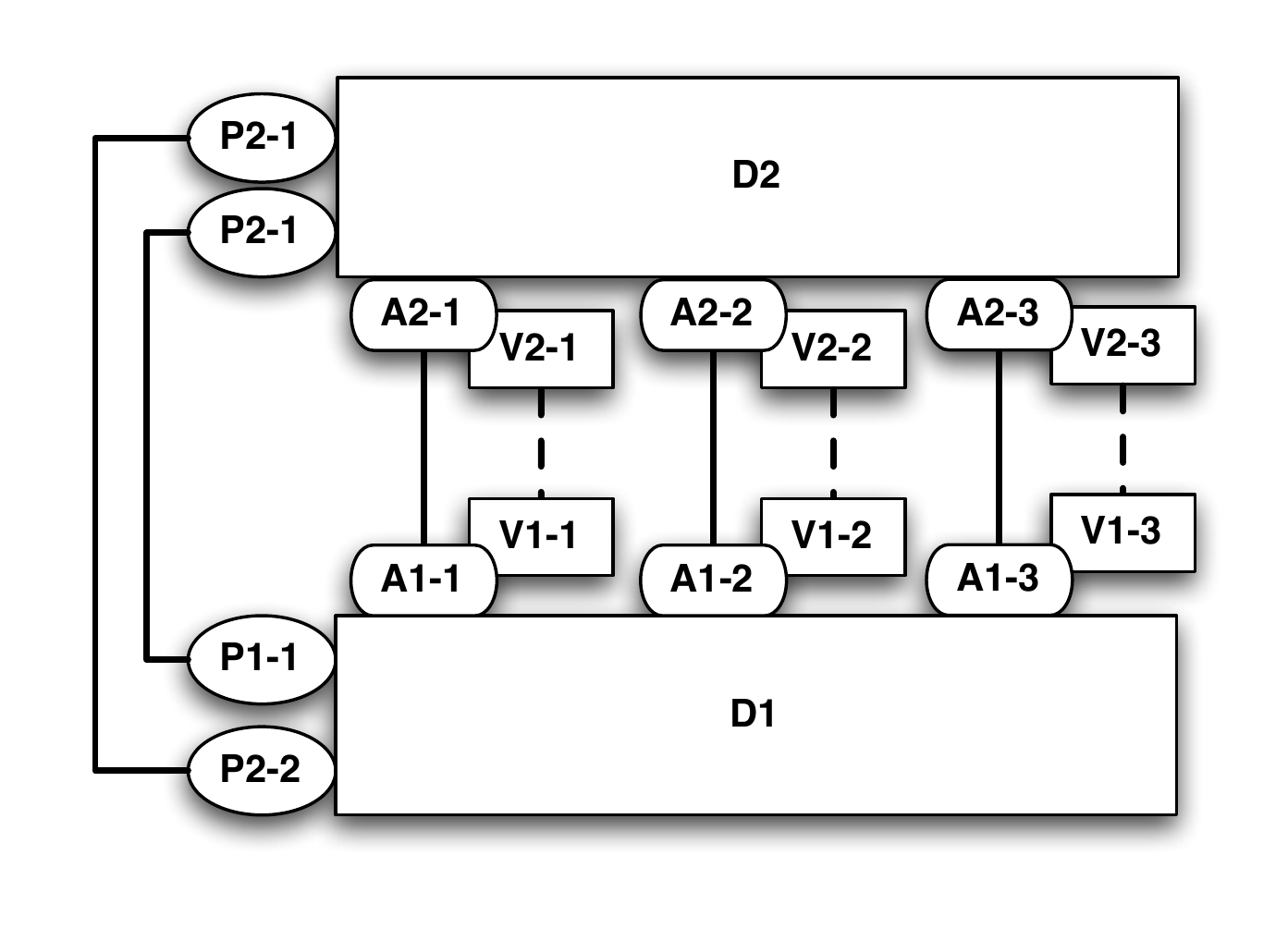}}
  \subfloat[Partitioned Dataset]{\label{fig:ex2}\includegraphics[width=0.3\textwidth]{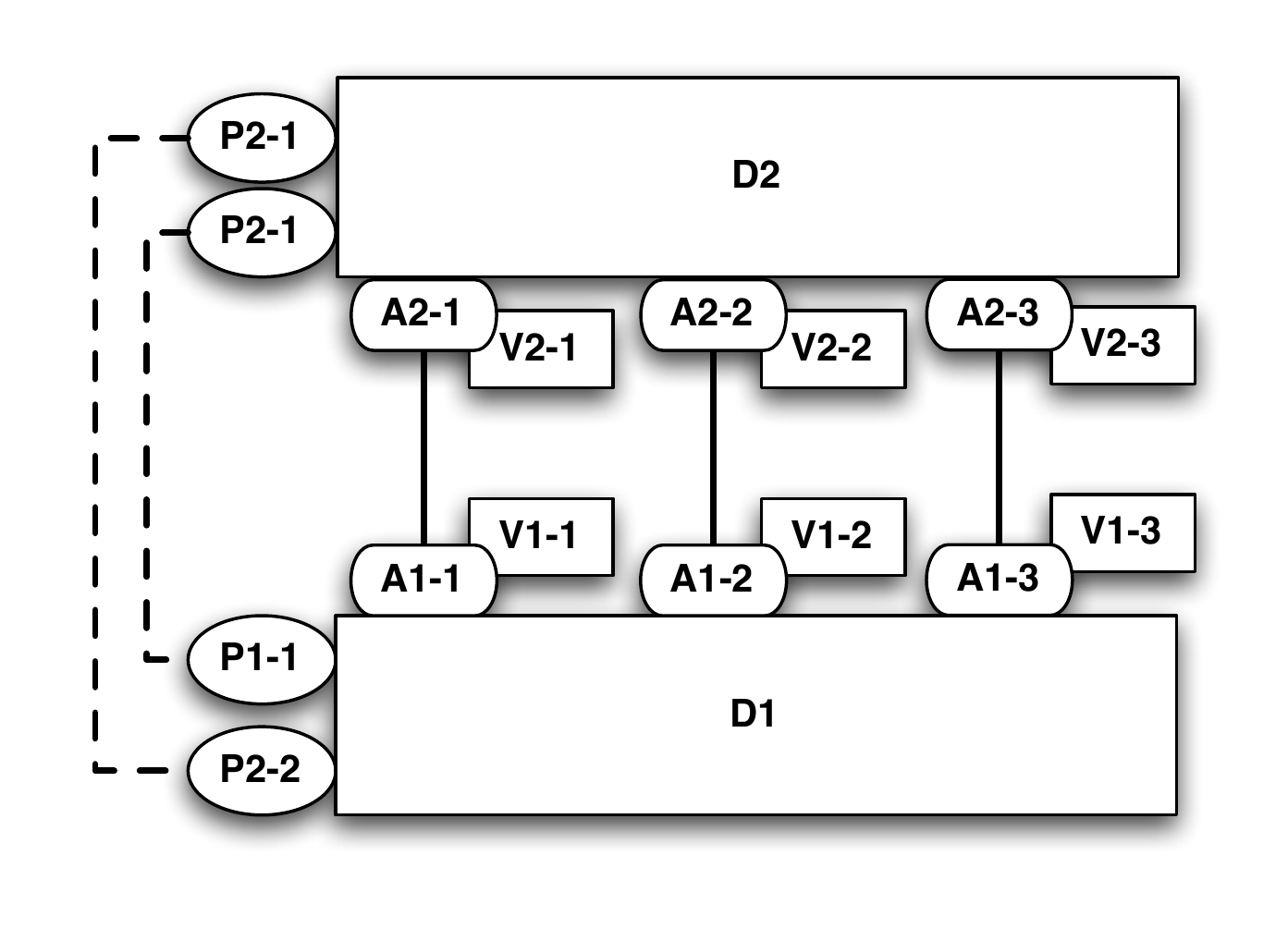}}
  \caption{Exemplary patterns leading to Error Correction Hypotheses}
  \vspace{-2mm}
  \label{fig:correction_hypotheses}
\end{figure*}
In addition, there is the \emph{matches}-relation, named $M$, which is the set of candidate matches between datasets.
Matches occur only between elements of the same type, and each match is assigned a confidence score.
Note that each element in $V$ represents the set of values for one attribute, not a single value.
This implies that connections between instances can only be expressed on a global level, i.e., as connections between two value sets.
The \emph{matches}-relation is the hypothetical part of the model, which can be instantiated by applying instance and schema matching techniques in a brute-force manner, comparing every dataset with every other dataset.
Note that it is not the goal of this step to produce the best possible mappings, as it is common in schema- or instance matching, but to generate integration hypotheses on the different levels of properties, attributes, and instances.\\
Another way to look at this model is as a graph of the platform's content, a notion which we will also use throughout the paper.
Figure \ref{fig:ex0} gives an impression of how a model for an exemplary Open Data Platform with four datasets might look like when represented as a graph.
The edges of the graph represent the elements of the \emph{matches}-relation, while the  different line types represent different scores assigned to the matches.
The next section shows how we use this content model to find and weight integration problems.

\section{Identifying and Weighting\\ Global Integration Problems} 
\label{sec:identifying_and_weighting_global_integration_problems}
With the content model instantiated for a given platform, integration hypotheses can be identified using pattern matching in the generated graph.
Formally, a hypothesis is defined as a relation $H \subseteq M$, with $H = H_{A} \cup H_{V} \cup H_{P}$.
Each hypothesis is of a certain class which is characterized by a specific configuration of edges, and is assigned a weight, whose calculation will be detailed in Section \ref{sub:weighting_integration_hypotheses}.\\
The hypothesis classes discussed in the following are not an exhaustive list of all integration problems that can be expressed in our model, but an exemplary subset.
In particular, we treat two main categories of hypothesis: \emph{error correction} and \emph{relation detection}.

\begin{figure*}[t]
  \centering
  \subfloat[Potential Join Partner]{\label{fig:ex4}\includegraphics[width=0.3\textwidth]{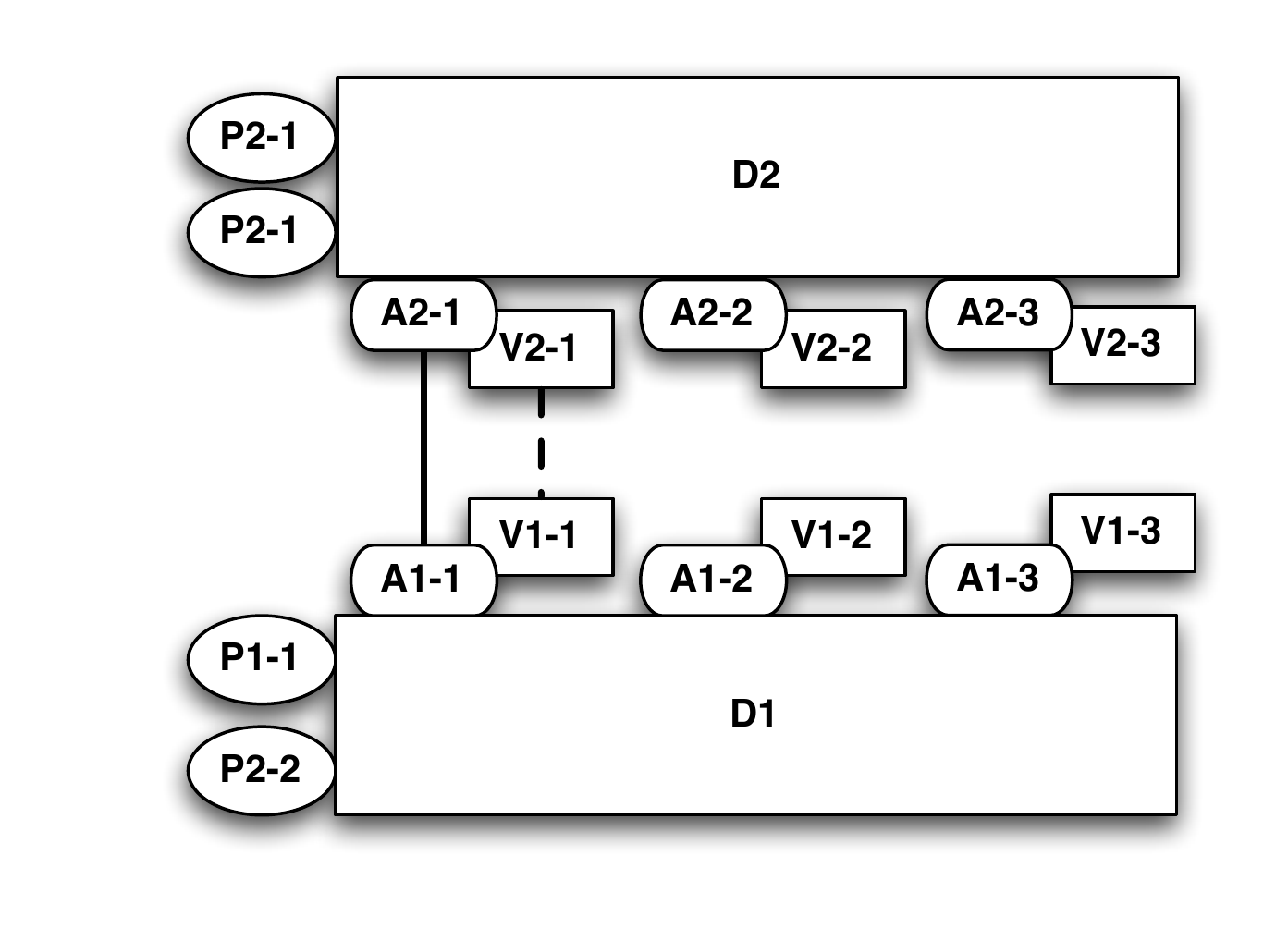}}
  \subfloat[Similar Domains]{\label{fig:ex5}\includegraphics[width=0.3\textwidth]{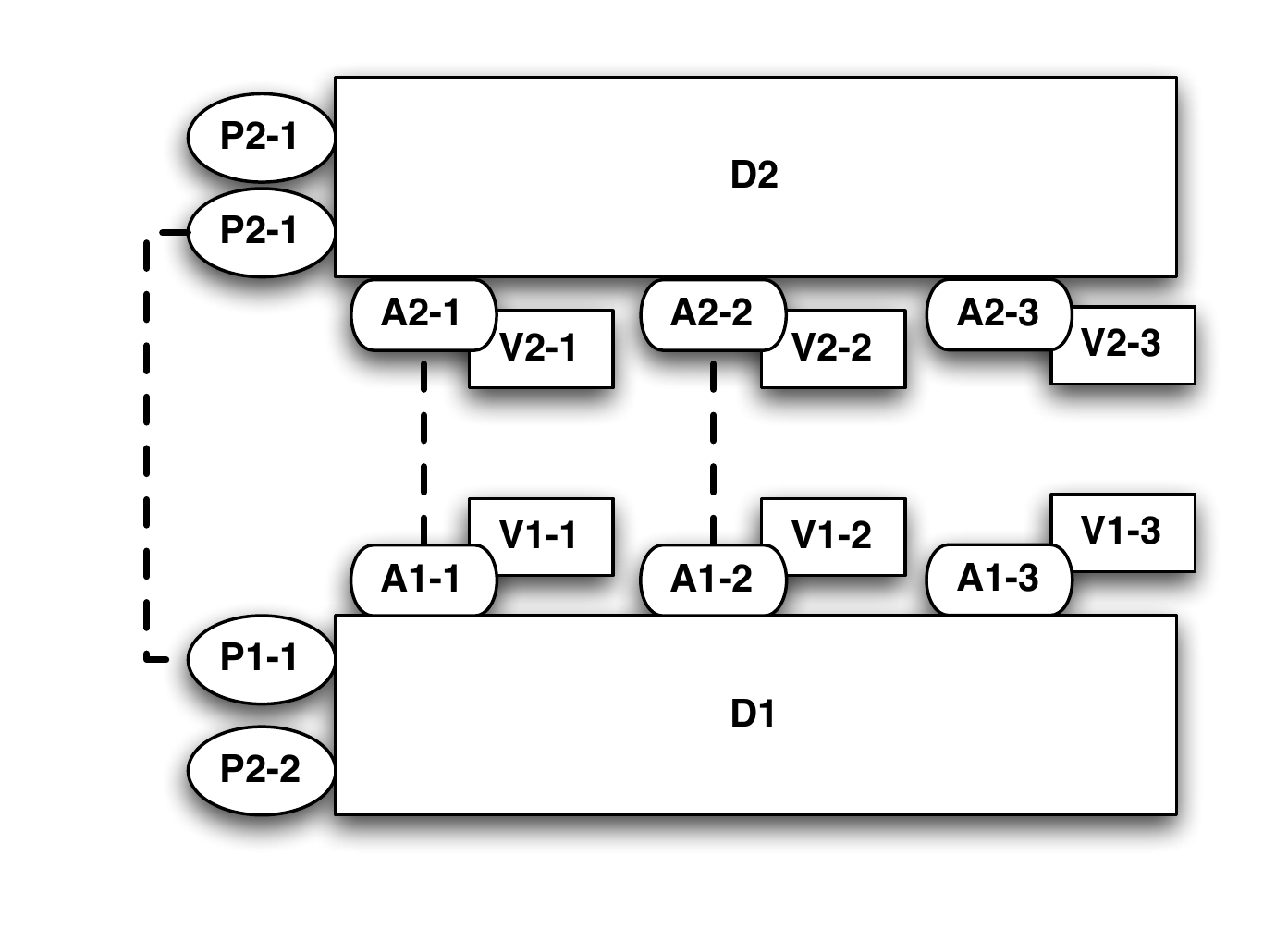}}
  \subfloat[Simple Relation]{\label{fig:ex6}\includegraphics[width=0.3\textwidth]{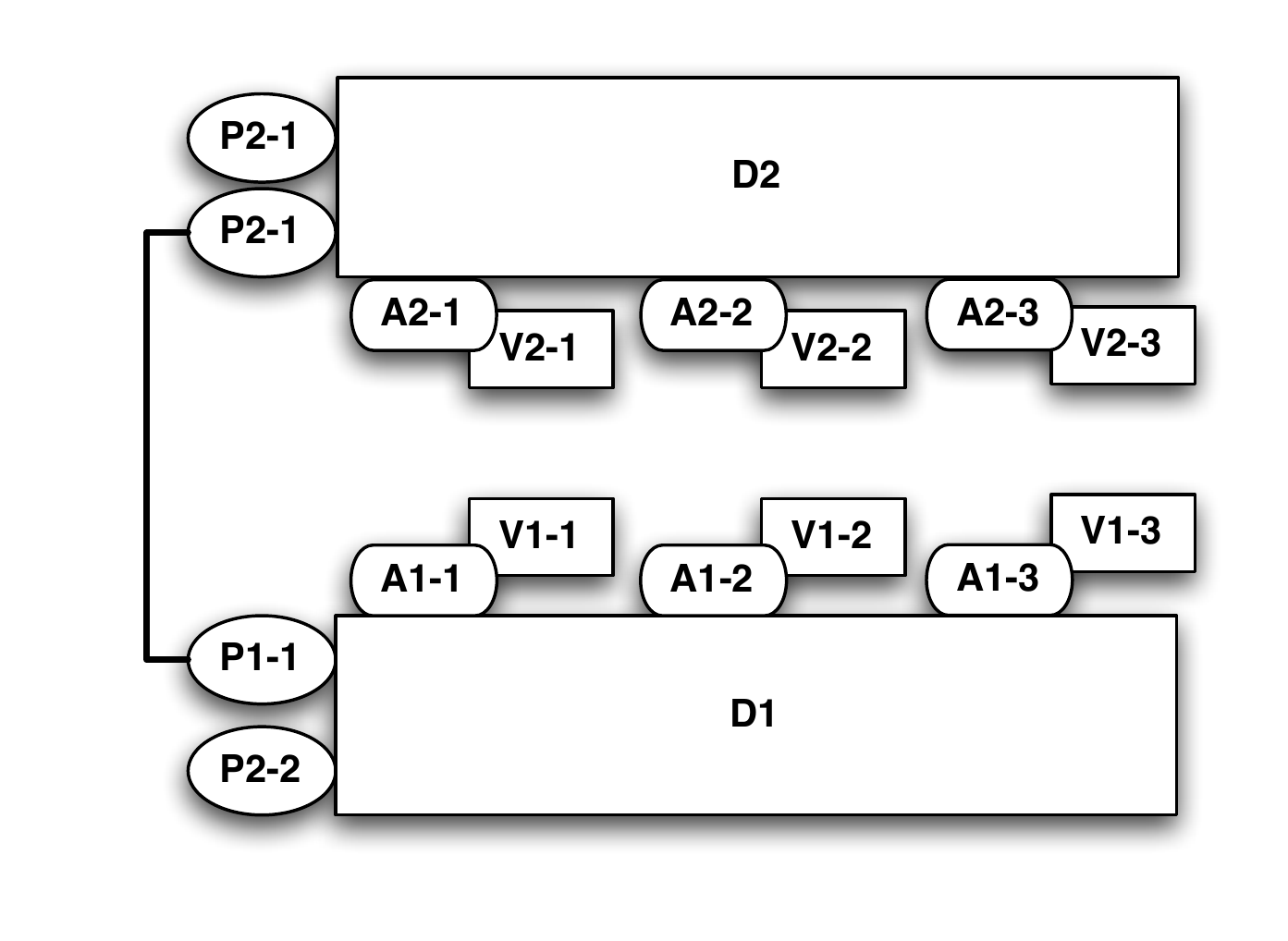}}
  \caption{Exemplary patterns leading to Correspondence Hypotheses}
  \label{fig:correspondence_hypotheses}
\end{figure*}

\subsection{Error Correction} 
\label{sub:error_correction}
While traditional integration problems are about bridging differences in vocabulary or structure, or more generally, resolving ambiguity, \emph{error corrections} problems are about removing actual errors from the content of a platform.
These problems stem from the free-for-all nature of Open Data Platforms as well as their lack of publishing standards and processes.\\
\textbf{Duplicate Datasets}: Figure \ref{fig:ex1} shows how a \emph{Duplicate Dataset} appears in the content graph.
In this example, two datasets match completely with their attributes as well as with their instance sets, while their metadata properties, such as title and description, show little similarity.
Intuitively, this could be caused by re-upload of the same dataset by different users of the platform, so the problem could solved by merging the metadata of both datasets and deleting one copy.\\
\textbf{Versioned Datasets}: Figure \ref{fig:ex3} shows a pattern that leads to a \emph{Versioned Dataset} hypothesis.
It occurs when modified versions of the same dataset are uploaded to one platform, e.g. when users publish their own views of common datasets or expand an existing dataset and republish it as a new dataset.
There can be changes in the instance set leading to sub- and supersets, as well as addition or removal of attributes leading to versions of one dataset.
The integration task in this case is to identify the largest dataset (instance- or attribute wise) if possible, or merge datasets if there is no clear best dataset.\\
\textbf{Partitioned Datasets}: The lack of publishing standards leads to datasets being uploaded to Open Data Platforms in several parts, i.e. a logical dataset is partitioned over a key and published as several physical datasets.
Examples include datasets being published periodically, e.g. for every year, or independently for several administrative areas.
This effectively corresponds to partitioning over a temporal or spatial attribute of one logical dataset.
Figure \ref{fig:ex2} shows how such a problem would manifest in the content graph.
The integration problem would be to create a new, explicit attribute, e.g. an attribute \emph{year} for a set of annually published datasets, and assign a value for each partition of the dataset.
\subsection{Correspondence Detection} 
\label{sub:correspondence_detection}
This category features problems where a potential semantic relation between datasets has to be verified or rejected.
They are more concerned with semantic correspondences and therefore more closely related to traditional data integration.
However, our goal is not to find mappings between schemata that model the same real-world entities, instead we aim to create ``semantic glue'' between quite different datasets to facilitate their reuse.
For example, many datasets on governmental Open Data Platforms are concerned with a certain state or county.
Even if they are unrelated in the domain they describe (e.g. education, energy, etc.), having explicit metadata about their common spatial attribute would be very beneficial.\\
\textbf{Potential Join Partners}: This type of integration hypothesis addresses the recombination of datasets from different sources.
A prerequisite for combining two datasets is finding attributes with common instances, that can interpreted as primary-/foreign keys in a join.
For example, two datasets about education spending and grades can be combined if a pair of common attributes can be found, e.g. about the county or school district.
Figure \ref{fig:ex4} depicts an example of how join candidates might appear in the content graph.
The integration task in this case would be to first verify the potential join partner (or discard it) and then verify and augment the mapping between the individual instances of the key attribute.\\
\textbf{Similar Domains}: Two different datasets that share a subset of their attributes can be thought of as having a similar domain, a fact that can be supported by overlap in their metadata properties.
For example, multiple datasets can originate from different sources and have quite different attributes, but if they share attributes such as ``amount'', ``beneficiary'' or ``receiver'', they could potentially all be concerned with government grants.
Figure \ref{fig:ex5} shows how this phenomenon is expressed in the content graph.
The integration task in this case would be to verify the connection and importantly, to label it, i.e. giving a name to the common subset of attributes.\\
\textbf{Simple Relation}: In the most general and simple case, a number of datasets can just have strong matches in a single metadata property, as shown in Figure \ref{fig:ex6}.
An example would be many datasets sharing a term in their metadata, e.g. ``2010 census'', which in this case implies that they all have a common origin.
The integration task in this case would be to verify whether there is a relation between the datasets, and possibly to name it.
\subsection{Weighting Integration Hypotheses} 
\label{sub:weighting_integration_hypotheses}
We have given a number of global integration hypothesis that can be identified on an Open Data Platforms using the content graph model.
Given limited resources, e.g., a limited amount of tasks that can be submitted to the crowd, it is apparent that ranking has to be performed.
We will now sketch our work-in-progress scheme for weighting them, which transforms a set of hypotheses $\mathcal{H}$ to an ordered list.
Intuitively, there are three factors influencing the relative weight, or importance, of an integration hypothesis: \emph{Probability of Verification}, \emph{Verification Cost} and expected \emph{Integration Benefit}.\\
Concerning the first point: Hypotheses can have different probabilities of being verified or discarded.
There are two possible strategies for incorporating these probabilities into the ranking:
\emph{most likely first} or \emph{most uncertain first}.
The first strategy emphasizes prioritizing hypotheses that are most likely to be verified.
For example, in the case of a join candidate, if the similarity between attribute names as well as instance set derived with automatic matching techniques is very high, it is more likely that the hypothesis will be verified by the crowd.
The second strategy would prioritize the hypotheses where the results of the automatic matching are most ambiguous, i.e. where contradictory indicators are found would be ranked higher.
Continuing the example, if the instances of two value sets match strongly, but there is no correspondence between attribute names in the respective datasets, this hypothesis would be favored, as it is more ambiguous.
Generally speaking, the \emph{most likely} strategy will result in more positive verification result, but might waste resources on easy problems where the automatic solution might have been acceptable.
The \emph{most uncertain} strategy will result in a higher information gain, because the harder problems are tackled, but might result in less positive integration results overall.
Formally, we define the probability of a positive verification as the average of the confidence values of all involved edges, regardless of type:
\begin{equation}
  p(H) = \frac{\sum_H c}
              {|H|}
\end{equation}
Note that this value corresponds to the \emph{most-likely} strategy.
For the \emph{most uncertain} strategy, the value should be normalized so that values close to the 0.5 are favored, assuming 0.5 implies highest uncertainty.\\
The second ranking factor, the cost of verification, is less straight-forward.
This cost varies between hypothesis types and specific occurrences.
In the crowd integration scenario, it can be measured by the number of tasks that have to be submitted to the crowd.
As an approximation, we use the number of edges in the hypothesis as this represents the number of matches that have to be verified.
Since a correspondence can only be verified in a meaningful way when the user is also presented the context correspondence, we add the size of the datasets, i.e. the context of the hypothesis, into the calculation.
\begin{equation}
  v(H_{d_1,d_2}) = \alpha * |H| + (1-\alpha) * (|d_1| + |d_2|)
\end{equation}
The weighting factor $\alpha$ between the size of the hypothesis and the size of the context can be learned by observing the actual integration results for different sized problems.\\
Of course, these costs have to be considered together with the third factor, the expected integration benefit.
Most of the literature does not consider the benefit of an integration task, as the integration problems themselves are usually taken for granted (for exceptions see the related work in Section \ref{sec:related_work}).
In the context of public data platforms, there are several factors which can be used to determine integration benefit, which can be grouped into two classes: dataset \emph{intrinsic} and dataset \emph{extrinsic} factors.
Intrinsic factors represent the value of the dataset itself, e.g., quantity such as the number of attributes or rows or quality such as the lack of missing values.\\
Extrinsic factors describe the value of the dataset in terms of it usefulness to the community.
These could be derived both from the platforms query workload, e.g., from search logs, or from its social features such as click and download counts or comments.
Furthermore, datasets that are well connected to other datasets, either through verified relations or through hypotheses, are potentially more recombinable and thus should be favored for integration.
Our current approach focuses on the extrinsic factors: If $\mathcal{H}_{d}$ gives us the number of hypotheses that include $d$, and $\upsilon(d)$ gives us the relevance of a dataset on the platform as determined from search logs and click counts, then:
\begin{equation}
  b(H_{d_1,d_2}) = \mathcal{H}_{d_1 \wedge d_2} * \upsilon(d_1) * \upsilon(d_2)
\end{equation}

In the next section, we will apply the method we described to a real world Open Data platform.

\begin{figure}[t] \centering
\includegraphics[width=0.5\textwidth]{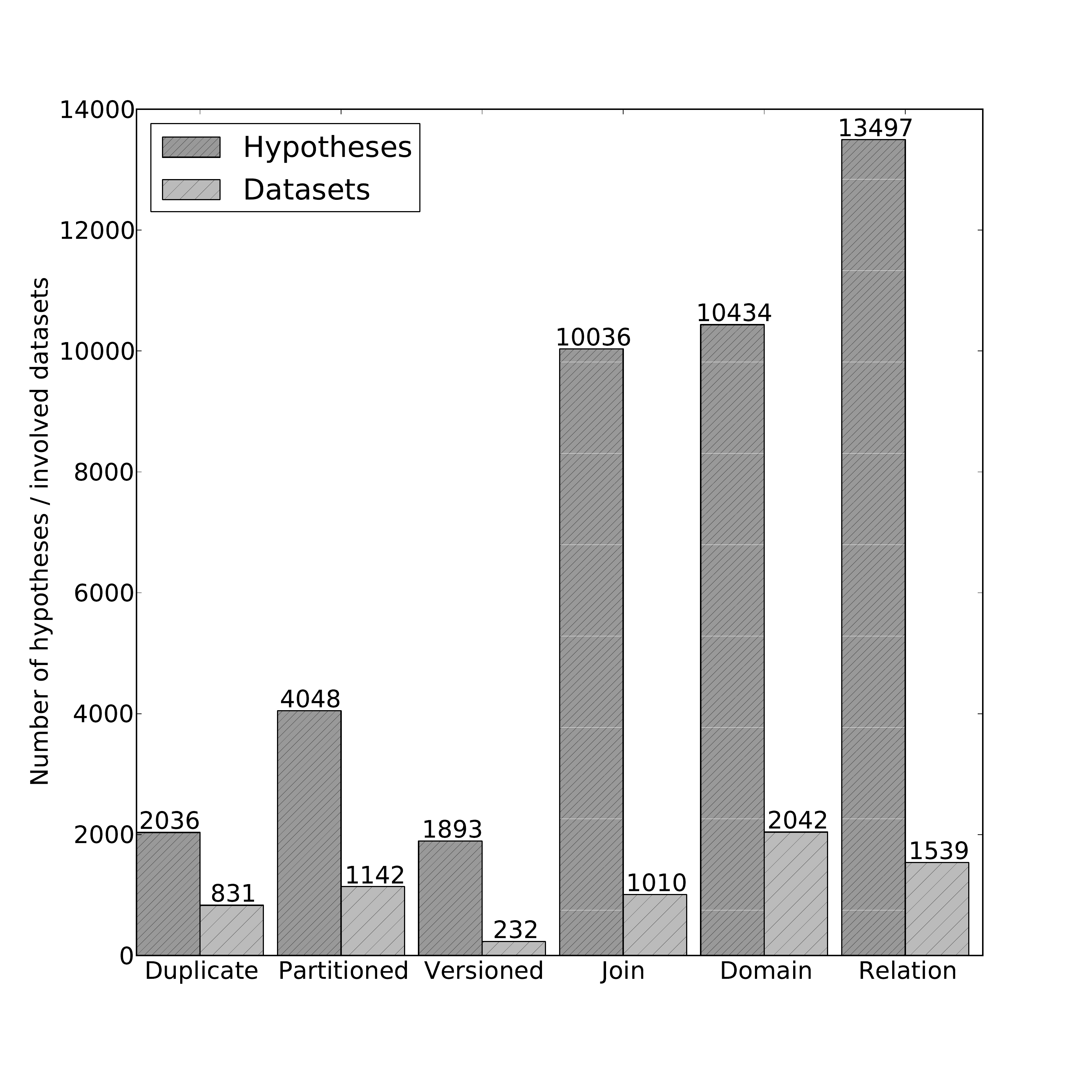}
  \vspace{-12mm}
\caption{Numbers of identified hypotheses and distinct datasets involved in them}
\label{fig:hypotheses}
\end{figure}

\section{Experimental Evaluation} 
\label{sec:experimental_evaluation}
For our experimental evaluation, we chose the well-known Open Data Platform \url{opendata.socrata.com}.
At the time of our experiments, the platform hosts about 18,000 individual datasets from various domains and origins, that can be accessed via a REST API, and that are presented in a relational format.
The platform is not bound to a specific agency or publisher (although it is operated by a commercial company), but instead allows any registered users to upload content.
The evaluation should verify whether our approach is able to (a) identify the six mentioned hypothesis types and (b) differentiate the relevance of the found hypotheses.\\
\begin{figure*}[ht]
\centering
\subfloat[Probability of Verification]{\label{fig:p}\includegraphics[width=0.25\textwidth]{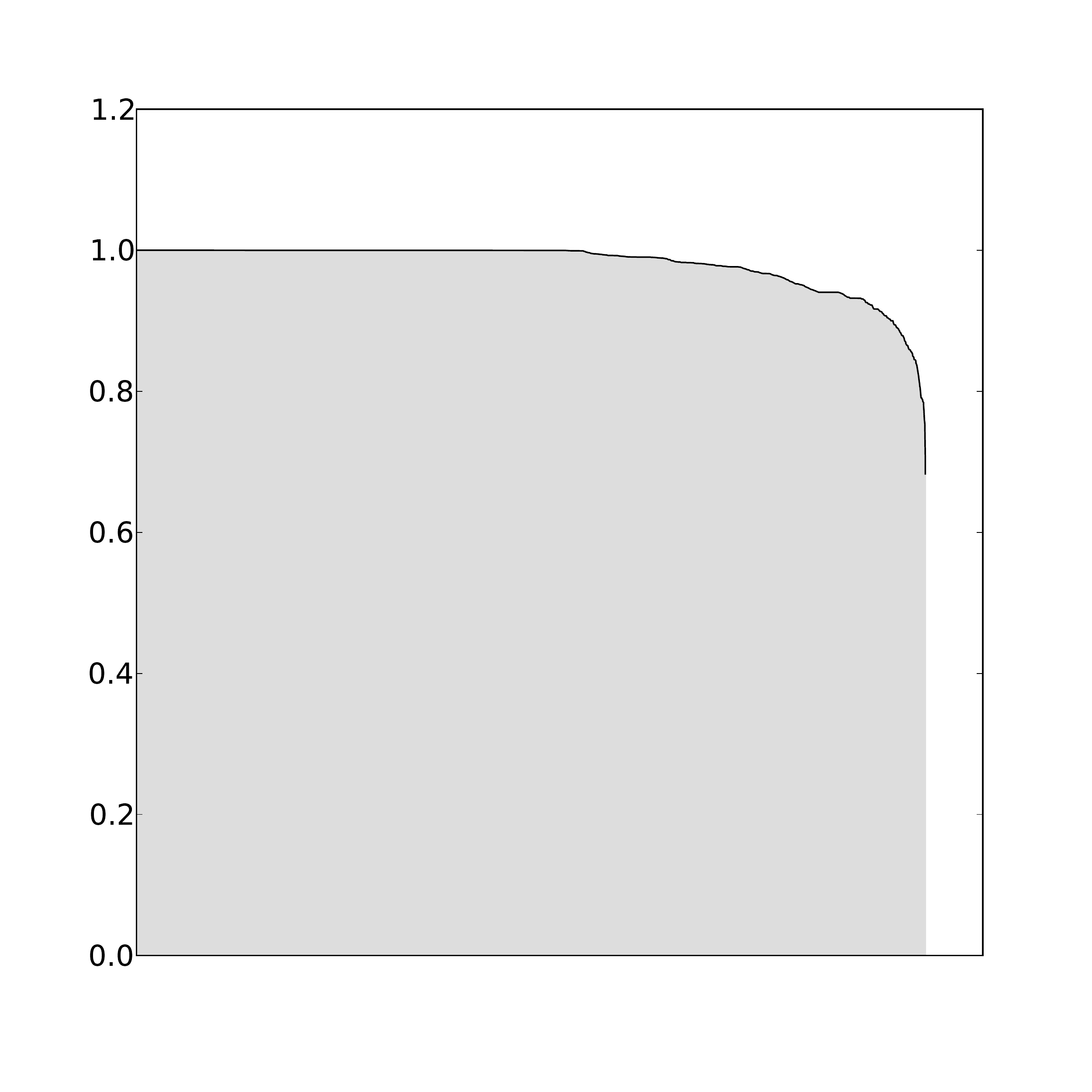}}
\subfloat[Cost of Verification]{\label{fig:c}\includegraphics[width=0.25\textwidth]{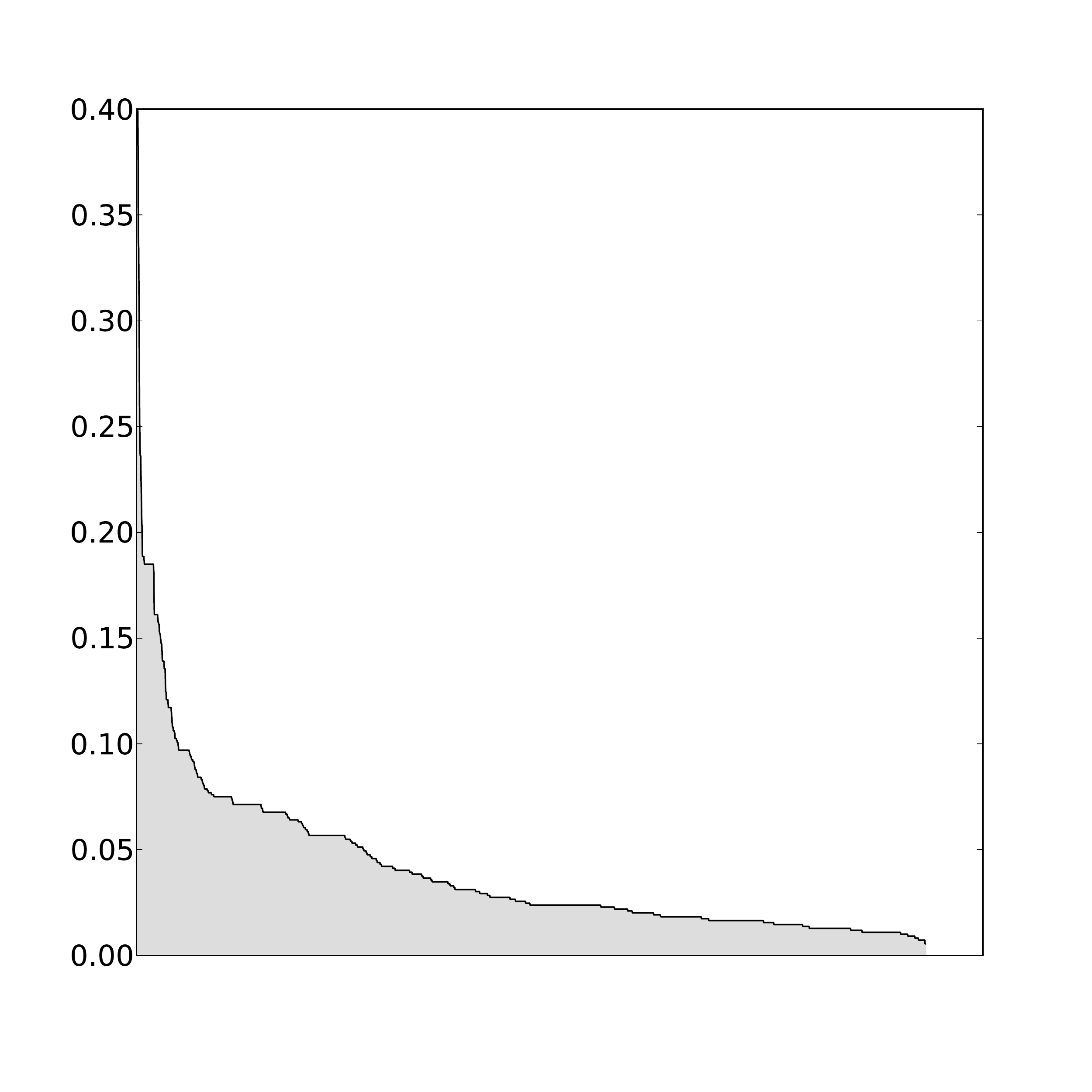}}
\subfloat[Integration Benefit]{\label{fig:b}\includegraphics[width=0.25\textwidth]{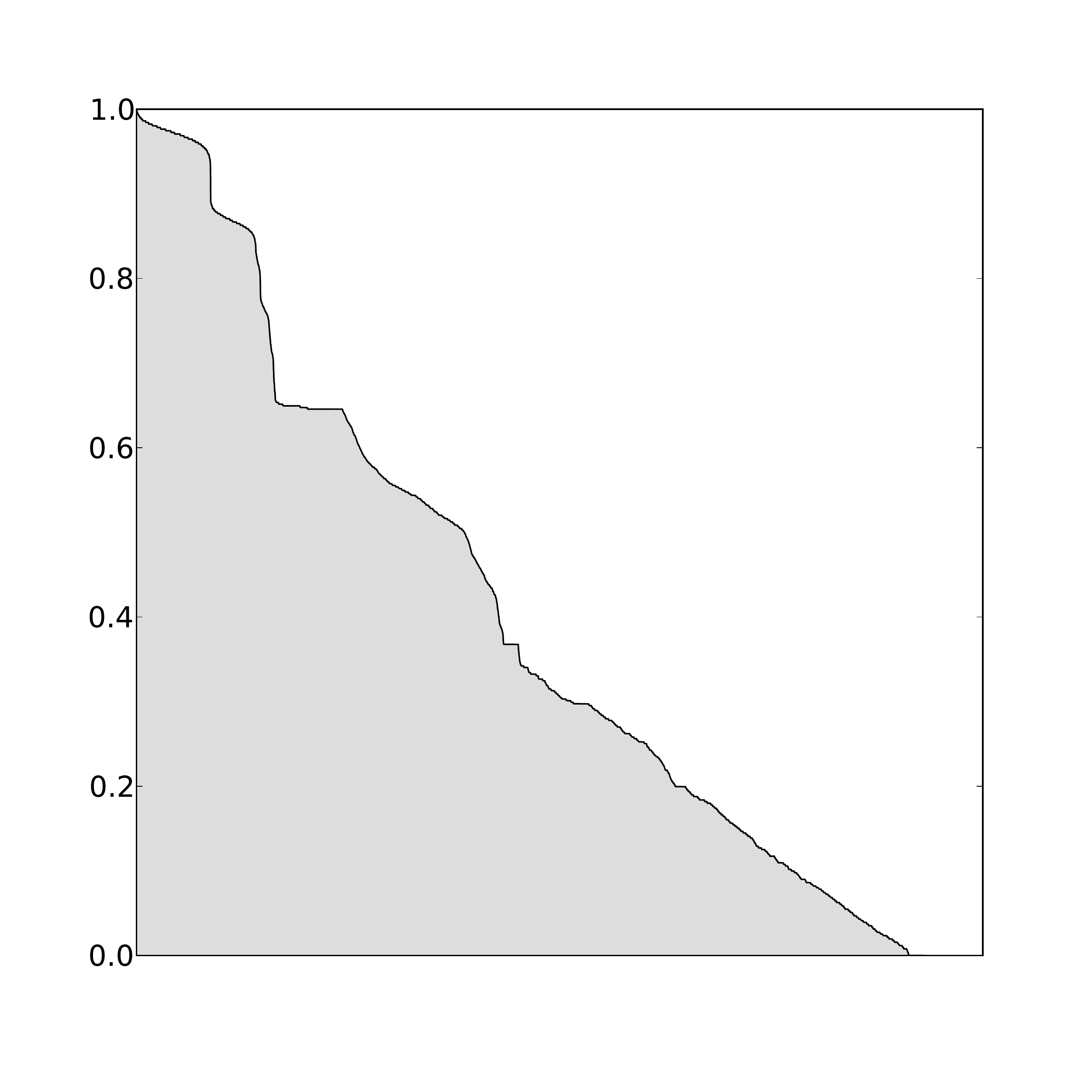}}
\subfloat[Ranking]{\label{fig:pcb}\includegraphics[width=0.25\textwidth]{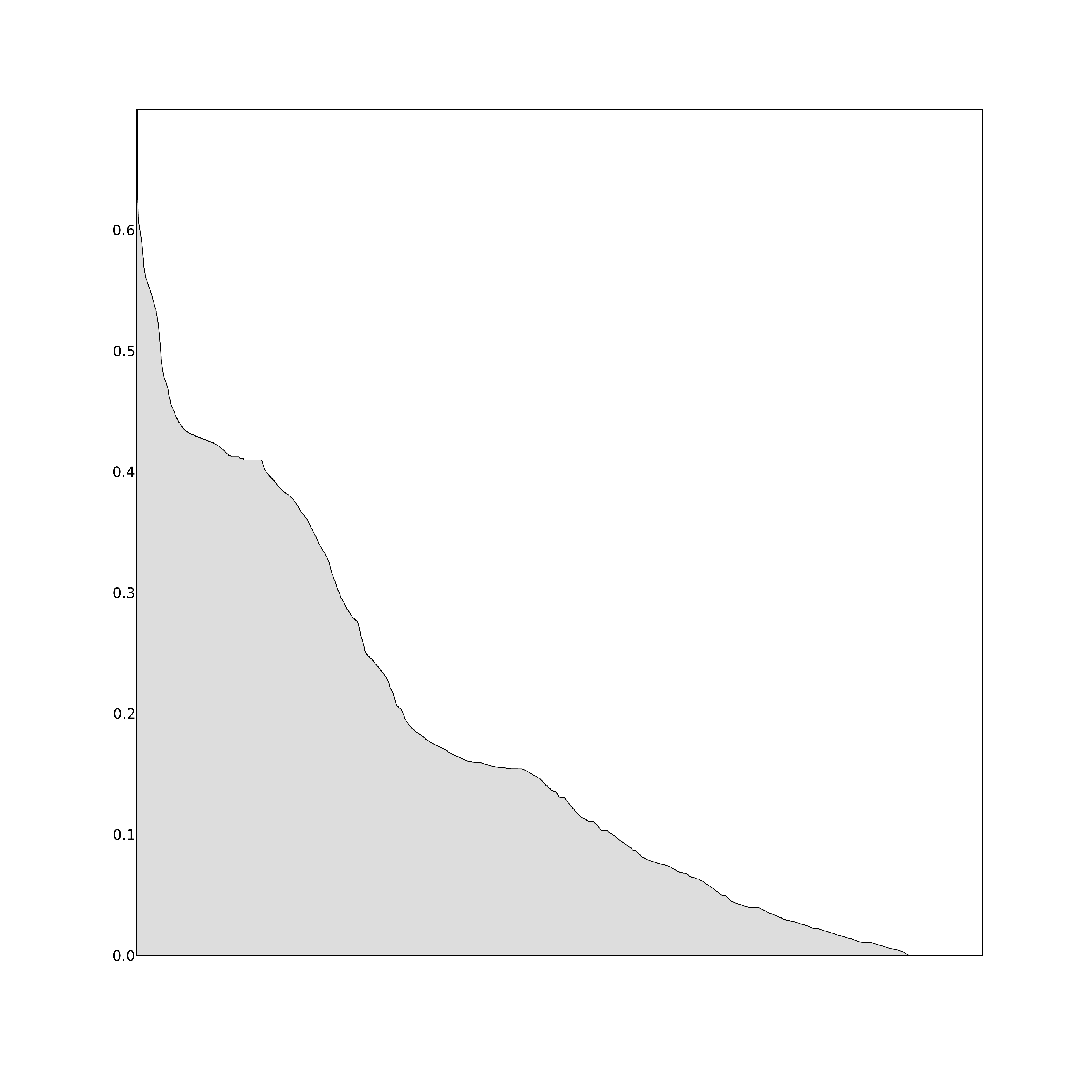}}
\caption{Distributions for probability, cost and benefit functions, normalized and ordered by rank}
\label{fig:pcbr}
\end{figure*}
For our experiments, we filtered very large datasets, datasets that contained no processable attributes and datasets with less than 50 rows bringing down the number of datasets to about 14,700.
With this, the number of datasets was reduced to about 3,700 with a about 15,000 attributes in total.\\
To create the content graph, we used a simple brute-force approach, matching every dataset against every dataset, comparing each individual attribute and its values to each from the other dataset.
In order to compare different attributes names we used simple string distance functions, such as the Jaccard and Levenshtein distances.
For properties, which include longer texts, we added TF/IDF-style weighting of the terms when matching.
For the instance sets, we matched the individual values and assigned the so called monogamy score of the created mapping, which indicates whether a one-to-one mapping between the two sets would be possible.
We generally used low thresholds for all similarity functions, as our method is about finding integration potential and thus about maximizing the recall and not about matching the precision.\\
We searched the resulting graph for the six patterns described in Sections \ref{sub:error_correction} and \ref{sub:correspondence_detection}.
Figure~\ref{fig:hypotheses} shows the number of hypotheses of the six examined types that where identified between the 3,700 datasets used in our evaluation.
Additionally, the number of distinct datasets involved in each of the hypotheses types is shown, e.g. 831 datasets were identified to have potential duplicates while there are 2,036 duplicate hypotheses.
This indicates that in average each dataset has between 2 and 3 potential duplicates on the Socrata platform.
In general, all six types of hypotheses could be identified in noteworthy numbers, confirming the need for global data integration methods on Open Data Platforms.\\
The results of applying our work-in-progress weighting schema to the Socrata data is shown in Figure \ref{fig:pcbr}.
The verification probability shows a large portion of matches with confidence value of 1.0 as determined by automatic matching.
Exact attribute name matches contribute the most to this plateau.
These datasets will receive no bonuses in the ranking, while datasets with less confidence are boosted.\\
In Figure \ref{fig:c} we see that costs on the other hand show a very small portion of the hypotheses with very high costs, and a long tail with relatively constant costs.
This tail represents the large share of single matches between datasets whose cost is only slightly affected by the size of their context.\\
The distribution of the benefit (see Figure \ref{fig:b}) shows several plateaus, which could be explained by large sets of partitioned or versioned datasets, which use the same terms and are connected to each other, leading to similar benefit scores for them.
Finally, the ranking in Figure \ref{fig:pcb} shows a small portion of hypotheses most favored for verification, followed by a relatively linear trend which are useful properties for a ranking function.\\
In summary, our method was able to automatically identify and weight integration potentials on a real world open data platform.
It can be considered as a first step towards a more integrated and thus more beneficial open data platforms.

\section{Related Work} 
\label{sec:related_work}
New forms of data management such as dataspaces and pay-as-you-go data integration~\cite{Franklin:2005,Madhavan:2007} are a hot topic in database research.
They are strongly related to Open Data Platforms in that they assume large sets of heterogeneous data sources lacking a global or mediated schemata, which still should be queried uniformly.
There is also an increasing number of works on incorporating human work directly into computations, for example about processing queries~\cite{Franklin:2011}, sorting and joining \cite{Marcus:2011}, or graph search~\cite{Parameswaran:2011a}, and even particularly for data integration using the crowd, such as in~\cite{McCann:2008,Hedeler:2011}.\\
Perhaps the most related paper to our work is \cite{Jeffery:2008} in which the authors solve a very similar problem for the generalized case of dataspaces.
They propose a novel method based on the \emph{value of perfect information} (VPI) to rank a set of candidate matches for verification in order of utility for the dataspace as a whole.
Similarly to our work, their ranking uses the confidence values of automatic matching methods combined with a measure for expected integration benefit.
In contrast to our work, their method is concerned with the most general integration task, matching of string pairs, while our approach treats specifics of Open Data Platforms on the level of datasets.
Furthermore, since they use atomic match candidates, the verification cost for every candidate is constant and therefore not considered, while our method assigns different costs for the verification of different hypotheses.
\balance

\section{Conclusion and Future Work} 
\label{sec:conclusion_and_future_work}
We have presented a work-in-progress method for identifying and weighting global integration problems in Open Data Platforms, that can be used to direct crowdsourced data integration or more generally, to find starting points for further integration methods.
We described several problem classes that appear on such platforms and how our method recognizes and scores them.
Finally, we applied the method to one of the largest Open Data Platforms and used it to find these problems in it, thus creating list of integration hypothesis for a real world setting.\\
For the future, we aim at creating integration techniques that combine crowdsourcing with traditional matching- and data mining algorithms that solves integration problems based on our generated hypotheses.
A different direction would be to construct a recommendation system for publishers, that performs the analyses presented in this paper when a new dataset is published, connecting it to the existing graph model of the platform.
Then, the system could generate feedback from the integration hypotheses it generated specifically for the new dataset.
This feedback could, for example, make the publisher aware of related datasets and the terminology they use, recommending attribute names for the new dataset that would allow easier discoverability on the platform and facilitate the combination with existing datasets.


\bibliographystyle{abbrv}
\bibliography{WOD2012}
\end{document}